\documentclass[conference]{IEEEtran}
\IEEEoverridecommandlockouts
\usepackage{cite}
\usepackage{amsmath,amssymb,amsfonts}
\usepackage{algorithmic}
\usepackage{graphicx}
\usepackage{textcomp}
\usepackage{xcolor}

\usepackage{cleveref}
\usepackage{fancyvrb}

\usepackage{booktabs}
\usepackage{multirow}
\usepackage{float}
\usepackage{svg}
\usepackage{makecell}
\usepackage{adjustbox}
\usepackage{enumitem}
\usepackage{xcolor}
\usepackage{soul} 
\usepackage[square,numbers]{natbib}

\usepackage{titlesec}
\titlespacing\section{0pt}{0.6\baselineskip}{0.3\baselineskip}
\titlespacing\subsection{0pt}{0.3\baselineskip}{0.15\baselineskip}
\titlespacing\subsubsection{0pt}{0.2\baselineskip}{0.1\baselineskip}

\def\BibTeX{{\rm B\kern-.05em{\sc i\kern-.025em b}\kern-.08em
    T\kern-.1667em\lower.7ex\hbox{E}\kern-.125emX}}

\makeatletter
\renewcommand\scriptsize{\@setfontsize\scriptsize{7.7}{8.7}}
\makeatother

\usepackage{fancyhdr,lipsum}
\fancypagestyle{firstpage}{
  \fancyhf{}
  \fancyhead[C]{To appear at the 3rd International Conference on Emergent Quantum Technologies (ICEQT'24), July 2024.}
  \fancyfoot[C]{\thepage}
}

\begin{document}

\title{Studying the Impact of Quantum-Specific Hyperparameters on Hybrid Quantum-Classical Neural Networks
\vspace{-10pt}
}

\author{\IEEEauthorblockN{Kamila Zaman\textsuperscript{1,2,*}, 
Tasnim Ahmed\textsuperscript{1,2,*},
Muhammad Kashif\textsuperscript{1,2},
Muhammad Abdullah Hanif\textsuperscript{1,2}, \\
Alberto Marchisio\textsuperscript{1,2}, 
and Muhammad Shafique\textsuperscript{1,2}}
\IEEEauthorblockA{\textsuperscript{1}eBrain Lab, Division of Engineering, New York University Abu Dhabi (NYUAD), Abu Dhabi, UAE \\
\textsuperscript{2}Center for Quantum and Topological Systems (CQTS), NYUAD Research Institute, NYUAD, Abu Dhabi, UAE\\
email: \{kz2137, tasnim.ahmed, mk9430, mh6117, alberto.marchisio, muhammad.shafique\}@nyu.edu
}
\thanks{*These authors contributed equally to this work.}
\vspace{-25pt}
}


\maketitle
\thispagestyle{firstpage}

\begin{abstract}
In current noisy intermediate-scale quantum devices, hybrid quantum-classical neural networks (HQNNs) represent a promising solution that combines the strengths of classical machine learning with quantum computing capabilities. Compared to classical deep neural networks (DNNs), HQNNs present an additional set of hyperparameters, which are specific to quantum circuits. These quantum-specific hyperparameters, such as quantum circuit depth, number of qubits, type of entanglement, number of shots, and measurement observables, can significantly impact the behavior of the HQNNs and their capabilities to learn the given task. In this paper, we investigate the impact of these variations on different HQNN models for image classification tasks, implemented on the PennyLane framework. We aim to uncover intuitive and counter-intuitive learning patterns of HQNN models within granular levels of controlled quantum perturbations, to form a sound basis for their correlation to accuracy and training time. The outcome of our study opens new avenues for designing efficient HQNN algorithms and builds a foundational base for comprehending and identifying tunable hyperparameters of HQNN models that can lead to useful design implementation and usage.
\end{abstract}

\begin{IEEEkeywords}
Quantum Machine Learning, Quantum Neural Networks, Quantum Hyperparameters.
\end{IEEEkeywords}

\section{Introduction}

Recent advances towards the development of quantum computers~\cite{Wu:2021,Madsen:2022} have opened doors to the exploration of many applications in the post-quantum era. One such application is \emph{Quantum Machine Learning} (QML)~\cite{Biamonte:2017,cerezo:2021,zaman2023survey}.
QML aims to benefit from the development of advanced ML models, architectures, training methodologies, and infrastructures to build ML algorithms, such as Deep Neural Networks (DNNs) in conjunction with the unique computational capabilities of quantum computers (QCs)~\cite{Huang:2021a,Kubler:2021,Abbas:2021}. 
Different QML algorithms have been analyzed for a variety of applications, including classification~\cite{schuld:2020,PrezSalinas:2020,blank:2020,Mari:2020,Benedetti:2019a,Coyle:2020, zaman2024comparative, innan2024fedqnn}. 
Inspired by the tremendous success of classical DNNs, their quantum counterparts \emph{Quantum Neural Networks} (QNNs) are one of the widely researched algorithms in QML.

However, the current state of QC technology is predominantly within the Noisy Intermediate-Scale Quantum (NISQ) era that is characterized by having a limited number of qubits, which are prone to errors, and lack of comprehensive error correction mechanisms~\cite{Preskill:2018}.
Parameterized quantum circuits (PQCs) have become a popular choice to design quantum algorithms in the NISQ era mainly because they are classically optimizable and provide robustness against device errors~\cite{schuld:2018}.
A QNN in general represents a PQC with encoded data~\cite{cerezo:2021,hubregtsen:2021}. However, as previously mentioned, the constraints inherent to NISQ devices make a purely quantum approach less practical at present. Consequently, many cutting-edge research efforts are focused on \emph{Hybrid Quantum Neural Networks} (HQNNs), which integrate both classical and quantum components~\cite{Bergholm:2018,Benedetti:2019,cerezo:2021,kashif:2021,Schetakis_2022SciRep_ReviewQML}.

HQNNs typically works in four primary steps~\cite{Kashif:2023}: (1) classical input preprocessing, (2) classical-to-quantum feature mapping, (3) training the PQCs on encoded data and measuring the qubits, and (4) classically postprocessing the measurement results of PQCs. 
HQNNs are designed to explore the potential advantages of integrating quantum components within a neural network framework~\cite{kashif:2021,kashif:2022a, maouaki2024advqunn}. 
This approach involves making a portion of the neural network \emph{quantum}, thereby combining the strengths of classical computing's robust data processing capabilities with QCs' unique computational power derived from quantum parallelism or quantum superposition.

\subsection{Target Research Problem}

Similar to classical DNNs, where the training process involves optimizing classical hyperparameters, HQNNs also require the optimization of quantum hyperparameters~\cite{Cong_2019, beer:2020}. This optimization is crucial in adapting the quantum components of the network to effectively process and learn from data. Since varying the parameters can lead to different circuit designs, utilizing hyperparameter tuning is essential for understanding the design space. It allows for laying the foundation to comprehend and design efficient quantum circuits. This paper focuses on contributing to the HQNN design space research as the main avenue of interest in the NISQ era, irrespective of benchmarking against state-of-the-art performance.
In HQNNs, the quantum hyperparameters typically relate to the settings of the \textbf{
number of layers \emph{a.k.a} depth of quantum circuit, type of entanglement, number of qubits, type of qubit measurement observable, and number of shots used to obtain the expectation value of qubits} in the underlying quantum layers that form part of the network.
These quantum hyperparameters determine how the qubits are manipulated during the QC process, impacting the overall behavior and effectiveness of the HQNN. 

\subsection{Research Gap}
The exploration of HQNNs for various ML applications has been a focal point in numerous studies. 
However, a significant portion of state-of-the-art research primarily focuses on the information carried by the number of qubits and/or the depth of quantum circuits as the key quantum parameters to assess the impact of these quantum modules on the computational complexity of these HQNN architectures. 

\textit{However, as previously discussed, there are several other quantum hyperparameters that can significantly influence the accuracy of HQNNs.} While classical hyperparameter tuning has been extensively studied, HQNNs introduce new parameters, i.e., quantum-specific hyperparameters, whose investigation is still relatively underexplored. 
To the best of our knowledge, there is no study that systematically examines the effects of a broad range of quantum hyperparameters on the overall accuracy and training time of HQNNs. Such a study would be valuable for understanding the full potential and limitations of HQNNs. It would provide insights into how various quantum hyperparameters interact and contribute to the accuracy of these networks, thereby guiding future circuit design advancements in the field of QML. Addressing this gap certainly leads to optimized and effective HQNN designs, tailored for specific applications, to leverage the strengths of both quantum and classical computing paradigms. 

\subsection{Contributions}
An overview of our novel contributions is shown in \Cref{fig:enter-label}. Following are their brief descriptions with key features: 

\begin{figure}[t]
    \centering
    \includegraphics[width=\linewidth]{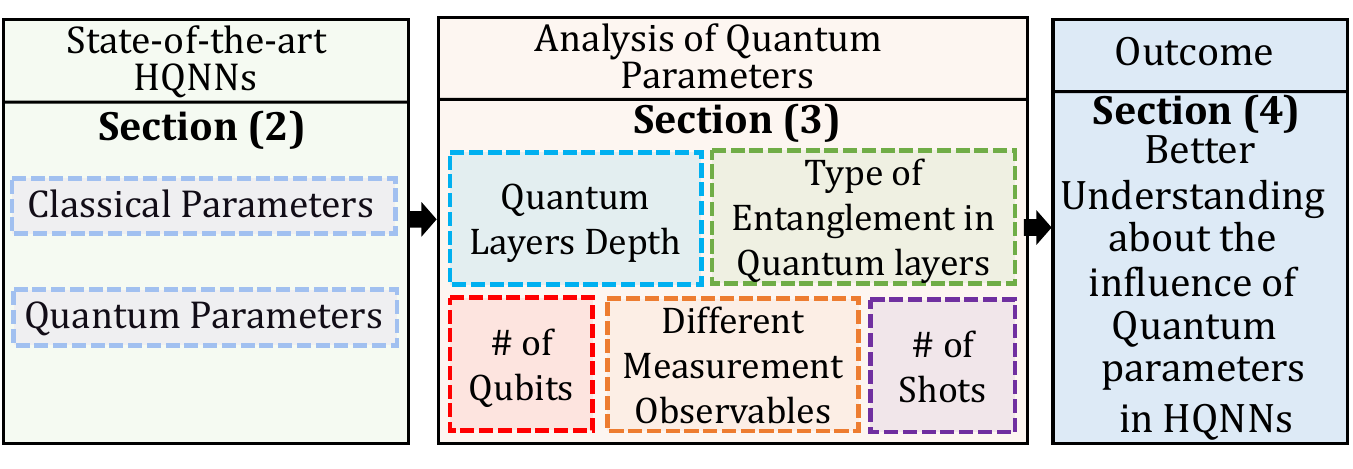}
    \caption{Overview of Our Contributions.}
    \label{fig:enter-label}
\end{figure}

\begin{itemize}[leftmargin=*]
    \item \textbf{Framework Design:} We have developed a comprehensive framework dedicated to evaluate the impact of various quantum-specific hyperparameters on the accuracy and training time of HQNNs. This framework provides a systematic approach for understanding the influence of these critical factors in HQNN circuit designs. Recording the accuracy and training time statistics for each experiment variation allows us to capture tradeoffs across each parameter setting.


    \item \textbf{Studying the Impact of Circuit Depth and Entanglement Variations:} We have explored the impact of different quantum circuit configurations in HQNNs. Specifically, our focus has been on circuits with varying entanglement strategies and depths, analyzing how these variations affect the overall accuracy and training time of HQNNs. 

    \item \textbf{Studying the Impact of Qubit Count:} We conducted a thorough investigation into the influence of the number of qubits on the training time and accuracy of HQNNs. This aspect of our research offers insights into how quantum resources impact the efficiency and accuracy of quantum-enhanced machine learning models.


    \item \textbf{Studying the Impact of Observables and Shots:} Finally, we have examined the role of qubit observables and the number of sampling shots in HQNNs. This investigation aims to discern their potential effects on the efficacy of these networks, thereby contributing to a deeper understanding of quantum measurement strategies in the context of quantum-enhanced machine learning.

\end{itemize}

%


\section{Background}
\subsection{QNN Architecture} \label{sec:QNNs_working}
\begin{figure}[t]
    \centering
    \includegraphics[width=\linewidth]{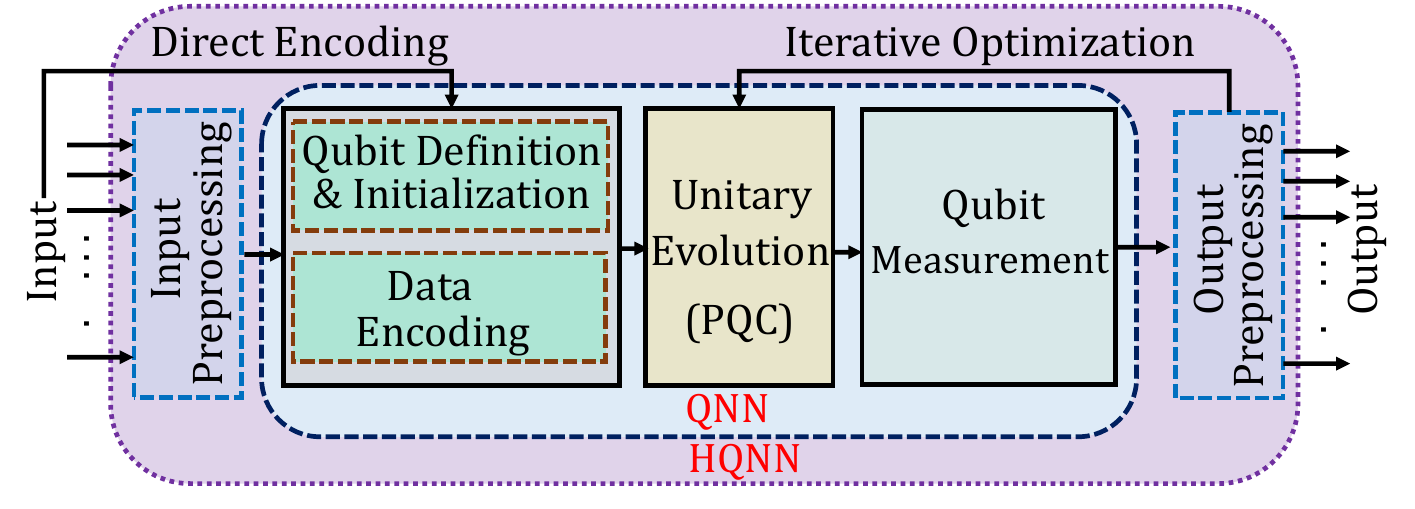}
    \caption{General Architecture of QNN and HQNN.}
    \label{fig:HQNN_arch}
\end{figure}

QNNs incorporate parameterized gate operations with learnable parameters~\cite{schuld:2020}. The typical workflow of QNNs, depicted in the blue-shaded region of Figure~\ref{fig:HQNN_arch}, starts with a qubit definition and initialization. A common convention for QNNs is to initialize the qubits in ground states. Once the qubits are prepared, the next step is the data encoding process, which is crucial for inputting the data into the quantum system. According to~\cite{Schuld:2021a}, the encoding step is generally static and not part of the training. However, there are proposals to make this routine trainable~\cite{Lloyd:2020}.
The embedded data is then processed via (trainable) PQCs and eventually measured to get the final output.  

\subsection{HQNN Architecture}
HQNNs leverage both classical and quantum computing resources for performing learning tasks, encompassing training and optimization processes. 
In HQNNs, a typical QNN architecture, as discussed previously, is employed, often supplemented with classical processing and/or post-processing 
techniques, as shown in Figure \ref {fig:HQNN_arch} (pink shaded region).
The preprocessing step involves downscaling the input features' size, to cope up with the limited number of available qubits in NISQ devices. Different techniques can be utilized for preprocessing step which includes, but not limited to, the use of classical layer with fewer neurons or some dimensionality reduction algorithm such as Principle Component Analysis (PCA)~\cite{kashif:2021,kashif:2022a}. 
All the input features are sequentially encoded by sliding the quanvolutional kernel over the input features~\cite{henderson2019quanvolutional}, and the encoded features are then passed to the QNN. The measurement results of the QNN are classically postprocessed, which is typically done through classical neuron layers and non-linear activations such as Softmax.

\begin{figure*}[h]
    \centering
    \includegraphics[width=.9\linewidth]{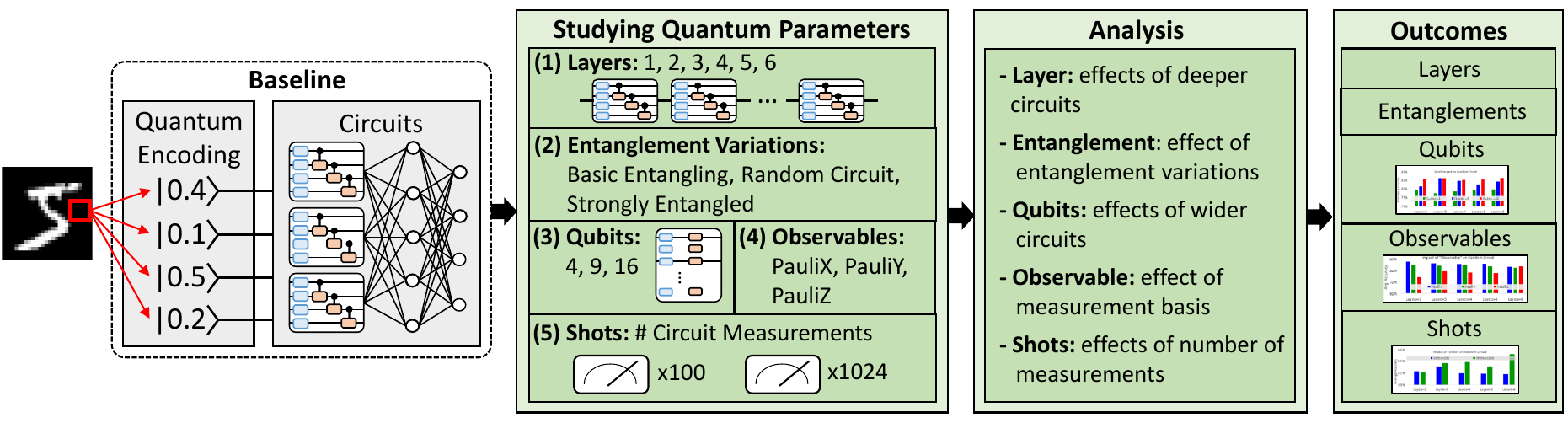}
    \caption{Overview of our comparative analysis methodology.}
    \label{fig:methodology}
\end{figure*}

\subsection{Data Encoding}
The feature mapping from classical data to quantum states is a crucial step in HQNNs, and can greatly influence their learning process. Different techniques are being explored to efficiently encode the data into a quantum system. However, the most frequently used encoding techniques in HQNNs are amplitude and angle encoding. In this work, we utilize angle encoding, which is widely adopted in the community as it is more suitable for HQNNs. It encodes the data into the rotation angles of qubits and can be described by Equation~\ref{eq1}. While using angle encoding, each qubit encodes a single input feature, hence, $n$ qubits will be required to encode $n$ input features.

\begin{align}
S_{{x}_{j}} = \bigotimes_{i=1}^{N} U_i \hspace{0.2cm} where \hspace{0.2cm}U_i &:=
    \begin{bmatrix}
 \cos(x_j^{(i)})  &  -\sin(x_j^{(i)})\\ 
 \sin(x_j^{(i)})   &\cos(x_j^{(i)}) 
\end{bmatrix} \label{eq1}
\end{align}

\begin{figure}[!t]
\centering
\includegraphics[width =\linewidth]{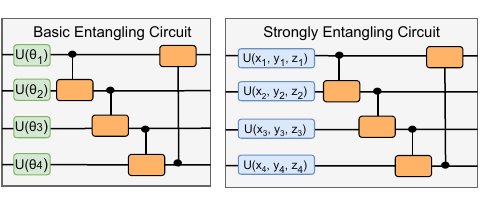}
\vspace*{-30pt}
\caption{Overview of (left) Basic Entangling and (right) Strongly Entangling circuits.}
\label{fig:basic}
\end{figure}












\section{Methodology}

We conduct a comprehensive analysis with the aim of better understanding the impact of quantum hyperparameters in Hybrid Quantum-Classical Neural Networks (HQNNs). To understand the contribution and convergence behaviors of the quantum hyperparameter variations, we focus on analyzing the correlation of a model's accuracy in relation to the learning curve attained over the training progress and model training time.
A detailed view of our methodology is depicted in \Cref{fig:methodology}. Below, we discuss different steps of our methodology in detail. 

\subsection{Layer Count Variation}

A circuit can be applied multiple times to a classical input. In our analysis, we refer to single quantum circuit in our architecture as a single layer. One of the focus of our experiment is to analyze the impact of circuit depth on model, which corresponds to repeating a layer multiple times. Instead of adding more gates to increase the depth of a circuit, we consider layer variations as one of our hyperparameters. By repeating a layer from 2 to 6 times before measurement, we study the impact of circuit depth in correlation with accuracy and training time.

\subsection{Entanglement Variation}

A single layer in our experiments can have three different entanglement orientation: Basic Entangling, Strongly Entangling and Random Circuit. 

    \textbf{\textit{Basic Entangling (BE) circuit}}, as shown in \Cref{fig:basic} (left), consists of one-parameter single-qubit rotation gates applied on each qubit, followed by ring of CNOT gates. In the ring of CNOT, each qubit is connected to its neighbouring qubit, with the last qubit considering the first qubit as the neighbour. 

    \textbf{\textit{Strongly Entangling (SE) circuit}}, as shown in \Cref{fig:basic} (right), it consists of three-parameter single qubit rotation gates on each qubit, followed by entanglers, which are defined in the PennyLane module, but the default is CNOT gate. 
    
    \textbf{\textit{Random Circuit (RC)}}, as the name suggest is generated from random subset of Pauli gates, single-parameter rotation gates, three-parameter single qubits rotations gates and entangling gates such as CX or CZ. Unlike, basic and strongly entangling, random circuit may or may not have entangling between the qubits, since the gates are picked at random. \textit{Note that, for the purpose of upholding determinism in our study, the random circuit generation has been fixed using a seed value. The fixed seed values allow every call within an experiment to generate the same random circuit, making it consistent over multiple layers in the architecture.}

\subsection{Qubit Count Variation}
The strength and capability of a circuit depends on the number of qubits it has. Therefore, in our experiments, qubit count is a hyperparameter that we vary in our circuit architecture to analyze how the width of a circuit impact the accuracy of a model with regards to its accuracy and execution time. We experiment with 4, 9 and 16 qubits circuit.

\subsection{Measurement Observable}
Observables are a measurable property of a quantum circuit, enabling us to measure a quantum state in different basis. A projective measurement, described by an observable M, is a Hermitian operator on the state space of the system being observed. Pauli Z is a frequently used basis for measuring circuits in QNNs. However, there are numerous other observables that can have an impact, which potentially affects the measurement outcome of a circuit and model accuracy. In our study, we experiment with measuring our circuits with Pauli X, Pauli Y, and Pauli Z observables because these three form a complete basis for the set of all unitary transformations on a quantum state. By changing the measurement basis of a circuit, we can observe the impact of different observables on the measurement outcome of a model.

\subsection{Number of shots}
Shots refer to the number of times a circuit is executed and measured. Multiple circuit execution outcomes are then used to calculate expectation and probability values of a circuit with respect to a basis. The higher the number of shots, the higher the HQNN accuracy is. Hence, we run our circuits with 100 and 1024 shots to better understand the importance and practical need of a higher number of shots as a tradeoff between accuracy and required compute cost (time).




\section{Experimental Results}

\subsection{Experimental Setup}
\label{notme}
In our experiments, given our choice of quantum hyperparameters, we have numerous circuit configurations for the quantum layer of our HQNN models. Hence, to ensure a fair comparison, we use a uniform classical optimization environment for both the classical and the quantum layers as specified in \Cref{tab:exp_setup}. In the HQNN architectures of our experiments, the quantum layer is followed by a single classical layer to convert the quantum measurement values into classical probabilities. As for the quantum layers, the PennyLane framework provides PyTorch integration modules, which convert a quantum layer into PyTorch trainable layers capable of performing classical optimization of the gate-based quantum circuits for training with hyperparameter variations specified in \Cref{tab:exp_variation}. 
In our experiments, to curb the high computing power required for quantum simulations, we use a subset of the MNIST dataset~\cite{Deng2012TheMD}. Thus, the last classical layer of our models consists of only 4 neurons (equal to the number of output classes).
 
\begin{table}[!t]
\centering
\caption{Training Environment Specifications}
\label{tab:exp_setup}
\begin{adjustbox}{max width=\linewidth}
\begin{tabular}{c|c}
\textbf{Algorithm} & \textbf{Experiment Name} \\ \toprule
Software Framework & \verb|PennyLane(PL)| \\ \midrule
Back-End Simulator & \verb| lightning.qubit (PL)|  \\ \midrule
Back-End Machine &\ \verb|NVIDIA RTX 6000 Ada|\\ \midrule
Deep Learning Interface & Pytorch \\ \midrule
Dataset & \verb|MNIST-digit|,  Classes \verb|0,1,2,3| \\ \midrule
Training Samples, Testing Samples & \verb|100|, \verb|100| \\ \midrule
Epoch, Batch Size, LR & \verb|5|, \verb|5|, \verb|0.01| \\ \midrule
Data Encoding & \verb|RY angle encoding| \\ \midrule
\end{tabular}%
\end{adjustbox}
\end{table}

\begin{figure}[!t]
\centering
\includegraphics[width =\linewidth]{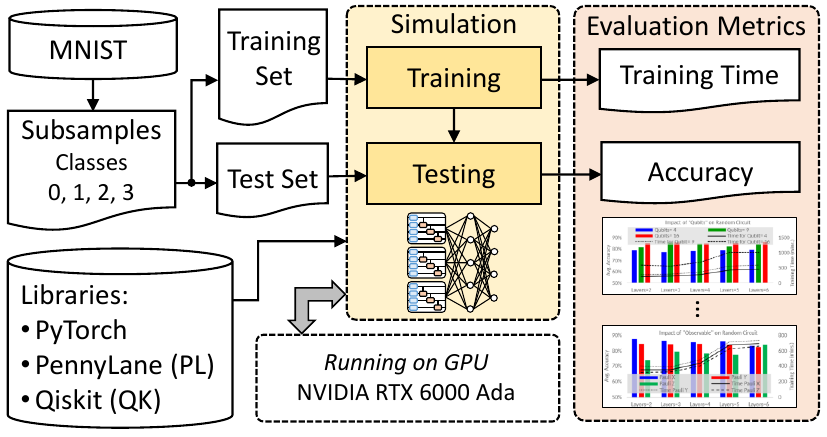}
\caption{Experimental setup.} 
\label{fig:experiments}
\end{figure}

\begin{table}[!ht]
\centering
\caption{Hyper-parameter Experiment Value Variations}
\label{tab:exp_variation}
\begin{adjustbox}{max width=\linewidth}
\begin{tabular}{c|c}
\textbf{Parameter} & \textbf{Tested Values} \\ \toprule
No. of Layers & \verb|2- 6|\\ \midrule
Circuits & \verb| Random Circuit|, \verb| Basic Entanglement| , \verb| Strong Entanglement|  \\ \midrule
No. of Qubits & \verb|4|, \verb|9|, \verb|16|  \\ \midrule
Observables & \verb|Pauli X|, \verb|Pauli Y|, \verb|Pauli Z|  \\ \midrule
Shots &\ \verb|100|, \verb|1024| \\ \midrule

\end{tabular}%
\end{adjustbox}
\end{table}

\subsection{Impact of Qubit Count vs. Circuit Variations}
\begin{figure}[!t]
\centering
\includegraphics[width =\linewidth]{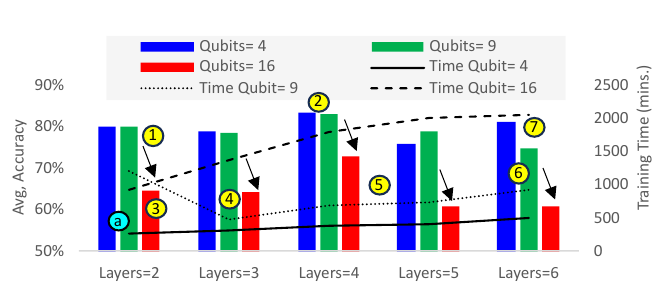}
\caption{Impact of Qubit Count on Random Circuit, derived by averaged performance value over shot variations (100 and 1024) and observable variations (Px, Py and Pz).} 
\label{fig: pl_qubit_impact_rc}
\end{figure}

\begin{figure}[!t]
\centering
\includegraphics[width =\linewidth]{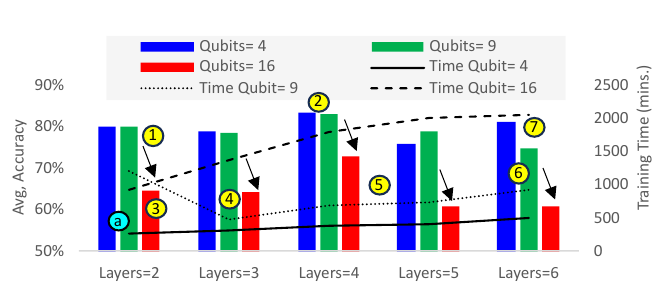}
\caption{ Impact of Qubit Count on Basic Entangling  derived by averaged performance value over shot variations (100 and 1024) and observable variations (Px, Py and Pz).} 
\label{fig: pl_qubit_impact_be}
\end{figure}

\begin{figure}[!t]
\centering
\includegraphics[width =\linewidth]{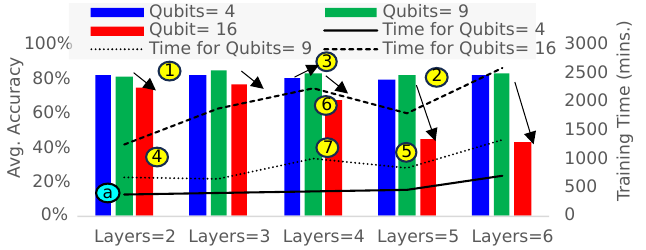}
\caption{ Impact of Qubit Count on Strongly Entangling  derived by averaged performance value over shot variations (100 and 1024) and observable variations (Px, Py and Pz).} 
\label{fig: pl_qubit_impact_se}
\end{figure}

\paragraph{\textbf{Random circuits}} The training results for random layers with different qubit count and circuit depths are shown in \Cref{fig: pl_qubit_impact_rc}. A clear pattern can be observed where increasing the number of qubits, irrespective of circuit depth, consistently results in better accuracy (see marker 1 in \Cref{fig: pl_qubit_impact_rc}). 
On the contrary, the training time drastically increases for an increased qubit count (see marker 2). 
Here, the training time gap between 4 qubits and 9 qubits is not as large as the increase in time from 9 to 16 qubits. 
Additionally, as indicated by markers 3, 4, and 5 in \Cref{fig: pl_qubit_impact_rc}, we notice that increasing layers within each qubit count variation circuit consistently increases the training time in a similar fashion across the 4, 9, and 16 qubit circuits. Despite the similar trend, when increasing the number of layers, the actual value difference in the training times across the variations of 4, 9, and 16 qubits is quite significant. As indicated by markers 2 and 6, we notice a significant distance between the time for 4 and 9 qubits and the 16-qubit circuit. However, this difference between them is proportional to the increased number of layers. This suggests that the capacity of a random quantum circuit to process information benefits from a higher qubit count but certainly comes with a significant training time cost. Marker a indicates the highest-accuracy experiment with random circuit is with $layers=6$ and $qubits =16$. 
From this observation, we can derive that for random circuit layers, deeper circuit with higher qubits are useful for better accuracy results. However, this is not the case for basic and strongly entangled circuits when compared to 16 qubit performances, as shown in Figures \ref{fig: pl_qubit_impact_be} and \ref{fig: pl_qubit_impact_se}.

\paragraph{\textbf{Basic Entangling Circuits}}  In contrast to random circuit results, for basic entangling layers, a rise in the number of qubits does not consistently enhance the accuracy, as shown by marker 1 in \Cref{fig: pl_qubit_impact_be}. 
While circuits with 4 and 9 qubits exhibit almost similar accuracy at shallow depths, increasing the depth to 6 layers reduces the accuracy for 9 qubits, as indicated by marker 7. This behavior can be attributed to the fact that 9-qubit circuits have relatively higher expressibility with deeper circuits, indicating a trade-off between circuit depth (i.e., number of layers) and width (i.e., number of qubits) to achieve reasonably better accuracy.
Moreover, circuits with 16 qubits show a marked decline in accuracy compared to 4 and 9-qubit configurations, regardless of the depth (see markers 3-6). This decline is attributed to the excessive expressibility of circuits with a large number of qubits that tend to approximate to a unitary 2-design. Hence, they are more prone to get trapped into barren plateaus, a phenomenon where gradient-based optimization becomes ineffective~\cite{McClean:2018}. 
The training time for 4 qubits is mostly consistent when increasing the depth (see marker a) and a slight increased time can be observed for 5 and 6 layers. 
However, we observe a significant training time increase while switching from 9 to 16 qubits (see markers 4, 5, and 6). 
Marker 2 highlights that, for the basic entangling circuit, the accuracy peak is reached with 4 layers for all qubit count values (4, 9, and 16).

\paragraph{\textbf{Strongly Entangling Circuits}} The results of strongly entangling circuits with different qubit count and circuit depth are shown in \Cref{fig: pl_qubit_impact_se}. The performance behavior in terms of accuracy is very similar to that of basic entangling circuits, i.e., as the number of qubits increases the accuracy starts to decline from 4 to 9 qubits and drops significantly when jumping from 9 to 16 qubits (see marker 1 in \Cref{fig: pl_qubit_impact_se}). Marker 3 highlights the accuracy peak for the strongly entangled circuits that, similarly to the basic entangled circuits, is achieved with 4 layers. 
The training time variation also follows similar patterns as the basic entangling circuit (see marker a), where a higher training time for 4 qubits can be appreciated when jumping from 5 to 6 layers. This behavior indicates that designing up to 5 entangling layers can be an efficient design decision. 
Marker 4 highlights a sharp training timeincrease when increasing the qubits. Markers 2, 5, 6, and 7 highlight a very interesting set of similar peaks and drops in time-cost across 9 and 16 qubits. In both cases, we observe a slight training time drop with 5 layers. 

\paragraph{\textbf{Training times of all circuits}} In terms of training time, all circuit types see increased training duration with greater circuit complexity. A notable observation is the relatively modest increase in training time when scaling from 4 to 9 qubits compared to the exponential increase from 9 to 16 qubits. This drastic rise highlights the computational challenges classical machines face in simulating larger quantum systems.
Comparing different circuit types, it is evident that random layers with a higher qubit count (16 qubits) have significantly lower training times than both basic and strongly entangling layers. Leading to better training efficiency and higher accuracy, even with a larger number of qubits.

These findings underscore the need for careful consideration of the interplay between qubit count, circuit depth, and gate complexity when designing quantum circuits for better training performance.

\subsection{Impact of Qubits vs. Circuit Variations}

To investigate the impact of different observables on the overall accuracy of HQNNs, we carry out the experimentation with 4 qubit circuits for varying depths. Below we discuss the impact of the observables for different circuits used in this paper. 

\begin{figure}[!t]
\centering
\includegraphics[width =\linewidth]{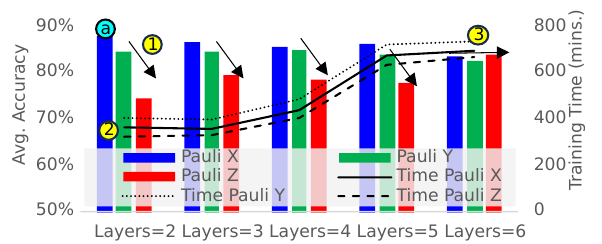}
\caption{ Impact of Observable on Random Circuit, derived by averaged performance value over qubit variations (4,9,16) and shot variations (100, 1024).} 
\label{fig: pl_observable_impact}
\end{figure}

\paragraph{\textbf{Random Circuits}} The experimental results for random circuits, as depicted in \Cref{fig: pl_observable_impact}, reveal that the choice of qubit observables have a notable effect on the accuracy of shallow-depth quantum circuits (with 2 to 5 layers). Specifically, when the qubits in underlying random quantum circuits are measured using the Pauli X observable, there is a marked improvement in the model's accuracy compared to measurements that employ Pauli X or Pauli Z observables, 
ss shown by markers a and 1 in \Cref{fig: pl_observable_impact}.
Marker 3 indicates that, for 6 layers, 
the accuracy becomes relatively uniform across the different observables. This could indicate that deeper circuits develop a level of complexity or expressibility that renders the specific choice of observable less impactful to the model's accuracy.

Marker 2 indicates the training time variation, with low differences across different varying observables. For all cases, while the time cost variation is minimal across the 3 obseravbles, we observe that Pauli Y always leads to the longest training time. 

\begin{figure}[!t]
\centering
\includegraphics[width =\linewidth]{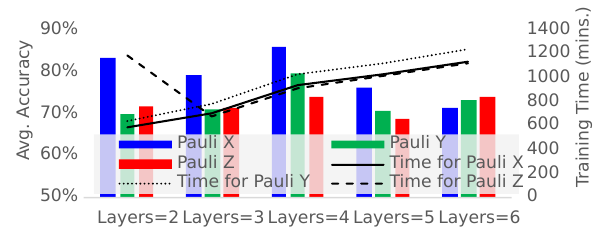}
\caption{ Impact of Observable on Basic Entangling Circuit, derived by averaged performance value over qubit variations (4,9,16) and shot variations (100, 1024).} 
\label{fig: pl_observable_impact_be}
\end{figure}

\begin{figure}[!t]
\centering
\includegraphics[width =\linewidth]{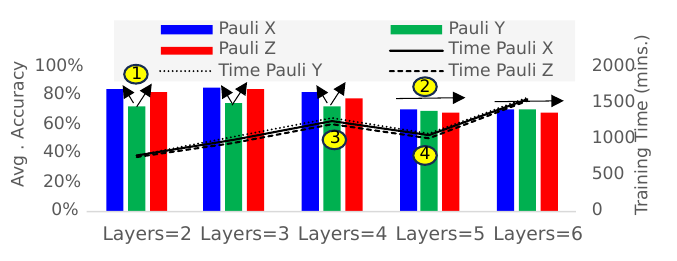}
\caption{ Impact of Observable on Strongly Entangling Circuit, derived by averaged performance value over qubit variations (4,9,16) and shot variations (100, 1024).} 
\label{fig: pl_observable_impact_se}
\end{figure}

\paragraph{\textbf{Basic and Strongly Entangling Circuits}} The empirical observations, as referenced in \Cref{fig: pl_observable_impact_be} and \Cref{fig: pl_observable_impact_se} for basic and strongly entangling layers regarding the influence of observables align with the findings in random circuits. In both circuit categories, the Pauli X observable demonstrates superior accuracy over others in circuits with shallow depth (see marker 1 in \Cref{fig: pl_observable_impact_se}). This superiority of Pauli-X diminishes or becomes insignificant with higher layers (see marker 2 in \Cref{fig: pl_observable_impact_se}). This stagnation shows a relatively lower learning capability for deep circuits. For less complex circuits, the choice of observable can be critical to the model's success, with Pauli X providing an edge in terms of accuracy. However, as circuit depth increases, the type of observable used becomes less impactful.

\paragraph{\textbf{Impact of Observables on training time}} The training for all the observables with different underlying circuit types used in this paper are shown in \Cref{fig: pl_observable_impact}, \Cref{fig: pl_observable_impact_be} and \Cref{fig: pl_observable_impact_se}. We notice that the choice of observable does not substantially influence the training duration. The training time remains relatively consistent across different observables, suggesting that the computational cost associated with training quantum circuits is largely independent of the observable type. On the other hand, markers 3 and 4 in \Cref{fig: pl_observable_impact_se} highlight a slight training time drop for 5 layers.

This uniformity in training time despite varying observables may stem from the fact that the computational cost of measuring different observables is not significantly different, or it is overshadowed by other, more dominant factors in the training process. Since observables are operators applied at the end of quantum circuit execution to extract information, their impact on the overall time complexity may be minimal compared to the time spent on preparing and processing the quantum states through the circuit layers. 
Moreover, this finding could imply that the choice of observable, while important for the accuracy, as previously discussed, does not significantly impact the training time. This is a valuable insight for the design of quantum algorithms, as it suggests that algorithm developers can select observables based on their suitability for the problem without concern for their impact on the training time.

     

 \subsection{Impact of Sampling Shots vs. Circuit Variations}
The sampling shots in a quantum circuit determine how many times a measurement is repeated to obtain statistical significance in the results. To assess the effect of the number of sampling shots on the accuracy of HQNNs, experiments were conducted using 4-qubit circuits across various circuit depths. This metric can have profound implications on the accuracy and reliability of quantum computations, particularly in the context of HQNNs, where classical and quantum components interact. Below we discuss the impact of shots on different types of circuits used in this paper. 

\begin{figure}[!t]
\centering
\includegraphics[width =\linewidth]{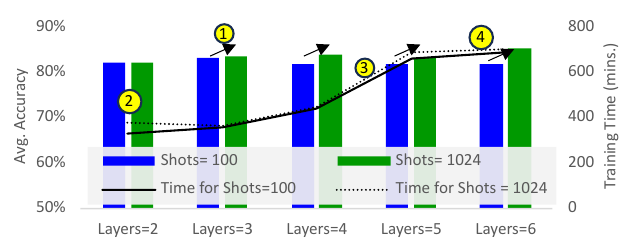}
\caption{ Impact of Sampling Shots on Random Circuit, derived by averaged performance value over qubit variations (4,9,16) and observable variations (Px, Py and Pz).} 
\label{fig: pl_shots_impact}
\end{figure}

\paragraph{\textbf{Random Circuits}}
The experiments with random circuits indicate that the number of sampling shots, whether at a lower count of 100 or a higher count of 1024, does not substantially affect the training performance of HQNNs in terms of accuracy. Despite higher shots results in a better accuracy (see marker 1 in \Cref{fig: pl_shots_impact}, their difference is minimal. Hence, for this setup, the choice of number of sampling shots does not influence the accuracy much. 
This consistency in accuracy across different shot counts suggests that, for random circuits, the accuracy of the trained model is relatively insensitive to the statistical fluctuations that might arise from the number of measurements.
Markers 2 and 3 show an increasing training time for both shot variations followed by stagnation when varying from 5 to 6 layers (see marker 4).

\begin{figure}[!t]
\centering
\includegraphics[width =\linewidth]{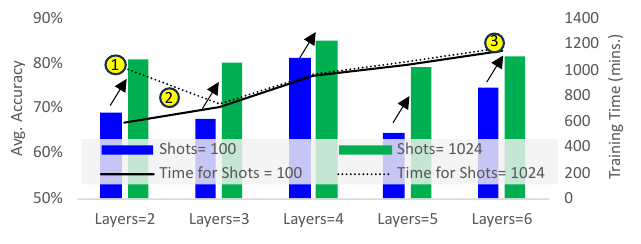}
\caption{ Impact of Sampling Shots on Basic Entangling Circuit, derived by averaged performance value over qubit variations (4,9,16) and observable variations (Px, Py and Pz).} 
\label{fig: pl_shots_impact_be}
\end{figure}
\begin{figure}[!t]
\centering
\includegraphics[width =\linewidth]{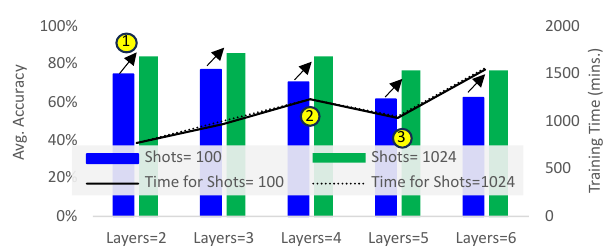}
\caption{ Impact of Sampling Shots on Strongly Entangling derived by averaged performance value over qubit variations (4,9,16) and observable variations (Px, Py and Pz)} 
\label{fig: pl_shots_impact_se}
\end{figure}

\paragraph{\textbf{Basic and Strongly Entangling Circuits}}
The experiments for basic and strongly entangling circuits, shown in \Cref{fig: pl_shots_impact_be} and \Cref{fig: pl_shots_impact_se}, respectively, show that a higher sampling shot count of 1024 leads to a noticeable improvement in accuracy across all circuit depths (see marker 1), with the relative difference being slightly higher for the strongly entangling circuit compared to the basic entangling circuit.
This marked difference suggests that, unlike for random circuits, the nature of basic and strongly entangling circuits makes them more sensitive to the number of sampling shots, likely because both basic and strongly entangling circuits incorporate more complex entanglement patterns. Higher sampling may be required to accurately capture the effects of these entanglements on the circuit's output. 

\paragraph{\textbf{Training time for all circuits}}
The training time for all types of quantum circuits used in the paper escalates with the increase in circuit depth. This trend can be attributed to their increased complexity and the subsequent computational demand of executing and optimizing deeper circuits. 
On the other hand, the impact of shots on the training time is insignificant (see marker 2 in \Cref{fig: pl_shots_impact_be}). 
Interestingly, the number of sampling shots, whether smaller (100) or larger (1024), does not induce a noticeable difference in training time. This could be because the major contributor to the training time may be the quantum processing itself, rather than the classical computation overhead associated with handling more shots.


\subsection{Results with Testing Data}

\begin{figure*}
\begin{tabular}{ccc}
  \includegraphics[width=55mm]{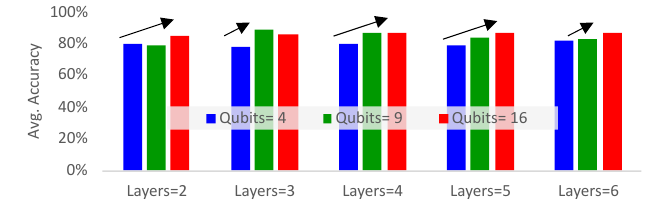} & 
  \includegraphics[width=55mm]{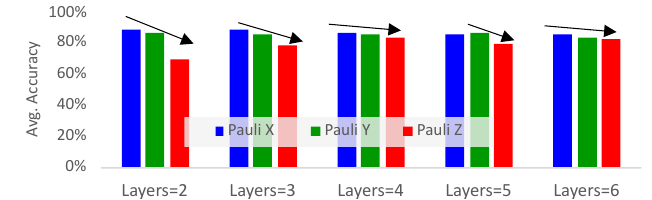} & \includegraphics[width=55mm]{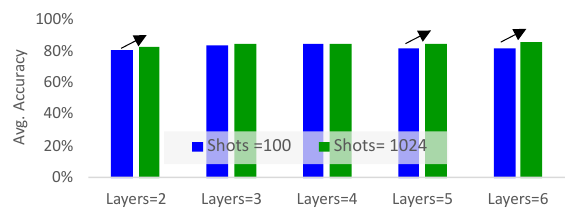} \\
(a) Random Circuit & (d) Random Circuit & (g) Random Circuit \\[1pt]
 \includegraphics[width=55mm]{images/pl_qubits_exp_test.pdf} & 
  \includegraphics[width=55mm]{images/pl_observable_exp_test.pdf} & \includegraphics[width=55mm]{images/pl_shots_exp_test.pdf} \\
(b) Basic Ent. Circuit  & (e) Basic Ent. Circuit & (h) Basic Ent. Circuit\\[1pt]
 \includegraphics[width=55mm]{images/pl_qubits_exp_test.pdf} & 
  \includegraphics[width=55mm]{images/pl_observable_exp_test.pdf} & \includegraphics[width=55mm]{images/pl_shots_exp_test.pdf} \\
  (c) Strongly Ent. Circuit  & (f) Strongly Ent. Circuit & (i) Strongly Ent. Circuit \\[1pt]
\end{tabular}
\caption{Test-Data accuracy results. (a), (b) and (c) are qubit count impact on all circuit types.(d), (e) and (f) are observable's impact on all circuit types and (g), (h) and (i) are shot's impact on all circuit types. When not specified otherwise, the experiments are derived by averaged performance values over qubit variations (4,9,16), observable variations (Px, Py and Pz), and shot variations (100, 1024).}
\label{fig:test_results_grid}
\end{figure*}

Fig. \ref{fig:test_results_grid} is a collection of test-data accuracy results that follows a similar representation as the training-data results that were presented in the previous sections. From careful observations, we can clearly see the experiment models perform with decent accuracy. Additionally, we can also observe that in most cases, the patterns and trends of the results (indicated by arrows in \Cref{fig:test_results_grid} are similar to their training counter-parts in terms of the result patterns visible.

\subsection{Key Result Observations}

The configuration of quantum-specific hyperparameters that lead to the highest accuracy is reported in \Cref{tab:exp_outcome}. A quantum circuit with 3 or 4 layers provides the highest accuracy over most experiments with varying values for other parameters. The training time is directly proportional to the number of qubits. The higher the qubits, the higher the execution time with an almost exponential increase. No significant change in compute time can be observed with increasing layers and shots.

\begin{table}[!t]
\centering
\caption{Key HQNN design guidelines for high accuracy.}
\label{tab:exp_outcome}
\begin{adjustbox}{max width=\linewidth}
\begin{tabular}{c|c}
\multicolumn{2}{c}{\textbf{Best Performance Experiment Parameters} } \\ \toprule
Qubit Count & \verb|16| \\ \midrule
Shots & \verb|1024|  \\ \midrule
Layers &\ \verb|3|\\ \midrule
Entanglement & \verb|Random Circuit| \\ \midrule
Observable & \verb|Pauli X| \\ \midrule
Accuracy & \verb|95.3%| \\ \midrule
Time & \verb|708.57 minutes| \\ \midrule
\end{tabular}%
\end{adjustbox}
\end{table}

The number of shots is directly proportional to the accuracy, except for the experiments with the Random Circuit, where there is no significant effect dependent on the number of shots. The Pauli X observable outperforms other observables in all experiments. However, the margin by which it does so varies. The relation between observable and entanglement type is interesting. For basic entangling circuits that have RX rotation gates, the Pauli X observable works well. On the other hand, for strongly entangling circuits that have U3 gates, both Pauli X and Pauli Z observables have similar accuracy.

\begin{figure}[!t]
\centering
\includegraphics[width =\linewidth]{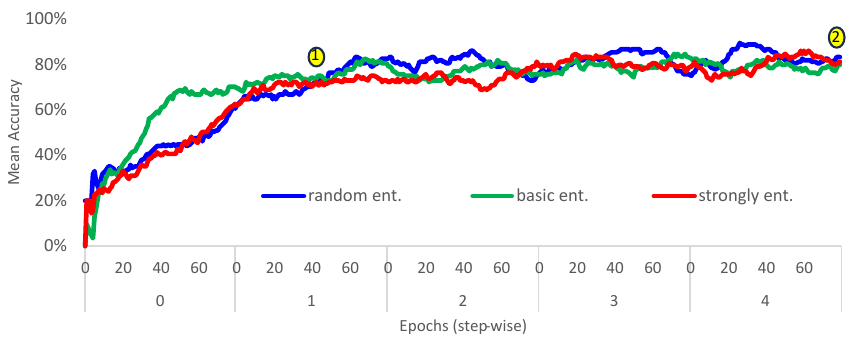}
\vspace*{-10pt}
\caption{Samples of epoch wise convergence in training accuracy for all three circuit variations, derived by averaged performance values over circuit depth (2, 3, 4, 5, and 6 layers) qubit variations (4,9,16), observable variations (Px, Py and Pz), and shot variations (100, 1024)..} 
\label{fig: epoch_convergence}
\end{figure}

\Cref{fig: epoch_convergence} illustrates the complete training process on step-wise granularity per epoch, for different entanglement variations. The results indicate a quick convergence towards a certain accuracy after a few training epochs. 
Indeed, the accuracy remains consistent within the last training steps (see markers 1 and 2 in \Cref{fig: epoch_convergence}), with no significant improvement when increasing the number of epochs. 


\section{Conclusion}

Compared to classical DNNs, HQNNs offer the possibility to tune not only the classical hyperparameters but also a set of quantum-specific hyperparameters that define the quantum circuit configurations. 
Such quantum-specific hyperparameters are the quantum layers depth, number of qubits, type of entanglement, type of encoding, number of shots, and measurement observables. 

In this paper, we comprehensively investigated the impact of these hyperparameters on HQNN models. We developed a systematic methodology to study the contribution of each hyperparameter, and their inter-dependency, with respect to the model accuracy and training time. Our findings advance the current understanding of HQNNs' functionality and behavior, paving the way toward efficient, robust, and accurate QML models. The observations derived from our experimental results uncovered trends and useful insights that allow QML designers to kickstart their experimentation environments by adopting efficient combinations of quantum hyperparameters. In future works, we plan to expand the experiment set with a wider variety of benchmarks, larger data samples, and conduct experiments on real quantum devices.



\section*{Acknowledgments}

This work was supported in part by the NYUAD Center for Quantum and Topological Systems (CQTS), funded by Tamkeen under the NYUAD Research Institute grant CG008.

\bibliographystyle{ieeetr}
\bibliography{main.bib}

\end{document}